\documentclass[12pt]{emulateapj}

\newcommand{\rd}{\mathrm{d}}
\newcommand{\rh}{\mathrm{h}}
\newcommand{\rf}{{\bf r}_0}
\newcommand{\rr}{{\bf r}}
\newcommand{\kk}{{\bf k}}
\newcommand{\trr}{\hat {\bf r}}
\newcommand{\nn}{{\bf n}}
\newcommand{\Psiot}{\tilde \Psi_{\mathrm{obs}} (\Omega, t)}
\newcommand{\Psiow}{\Psi_{\mathrm{obs}} (\Omega, \omega)}

\begin{document}

\title{Green's functions for far-side seismic images: a polar expansion approach}

\author{F. P\'erez Hern\'andez}
\affil{Instituto de Astrof\'\i sica de Canarias, E-38200 La Laguna, 
       Tenerife, Spain}
\affil{Departamento de Astrof\'\i sica, Universidad de La Laguna, E-38205 La 
       Laguna, Tenerife, Spain}
\email{fph@iac.es} 

\and 
\author{I. Gonz{\'a}lez Hern{\'a}ndez}
\affil{ National Solar Observatory\footnote
	{Operated by the Association of Universities for Research in 
	Astronomy, Inc. under cooperative agreement with the National 
	Science Foundation.},  950 N. Cherry Ave., Tucson, AZ, USA}

\begin{abstract}
We have computed seismic images of magnetic activity on the far surface of the Sun by using
a seismic-holography technique. As in previous works, the method is based on the 
comparison of waves going in and out of a particular point in the Sun but
we have computed here the Green's functions
from a spherical polar expansion of the adiabatic wave equations in the Cowling 
approximation instead of using the ray-path approximation previously used in the far-side
holography. 
A comparison between the results 
obtained using the ray theory and the spherical polar expansion is 
shown. We use the gravito-acoustic wave equation in the local plane-parallel limit 
in both cases and for the latter we take the asymptotic approximation for the 
radial dependencies of the Green's function.
As a result, improved images of the far-side can be obtained from
the polar-expansion approximation, especially when combining the Green's functions 
corresponding to two and three skips.
We also show that the phase corrections in the Green's 
functions due to the incorrect modeling of the uppermost layers of the Sun can
be estimated from the eigenfrequencies of the normal modes of oscillation.

\end{abstract}


\section{Introduction}

The idea that the observed wavefield at the solar surface can be used to estimate
the wavefield at any other location in the Sun has resulted in the development of several
local helioseismology methods such as the seismic-holography technique \citep{lb1997}.
Although the technique has been used to characterize flows 
in the solar subphotosphere \citep{braun2004} and to investigate seismic sources from flares 
\citep{donea2006}, its application to map areas of strong magnetic field \citep{lb1998} has 
remained a major focus. In particular, the mapping of active regions in the non-visible (or 
far side) of the Sun \citep{lb1990,lb2000a,gh2009} has proven to be a powerful tool for 
space weather forecasting.
A general review on local helioseismology can be found in \citet{gb2005}. 

The use of phase-sensitive seismic holography to map areas of strong magnetic field is based 
on the fact that there is a travel delay or phase shift between waves entering and exiting an
active region \citep{braun1992,duvall1996}. 
The technique compares theoretical waves going in and out 
of a particular point in the Sun (the focus) with the observed wavefield,
$\Psiot$, in some portion of the solar surface 
(the pupil $\mathcal{P}_\pm$), 
by computing the ingression and egression functions given by,
\begin{equation}
\tilde H_{\pm}(\rf, t_0)
= \int \rd t \int_{\mathcal{P}_\pm} \rd \Omega
\, \tilde \mathcal{G}_{\pm}(\rf, t_0 ; R,\Omega, t) 
\, \Psiot
\, ,
\label{eq:H}
\end{equation}
where $\rf$ is the location of the focus, $R$ the solar radius, $\Omega=(\theta,\varphi)$, 
$\tilde \mathcal{G}_+ (\rf, t_0 ; R,\Omega, t)$
is a Green's function that represents the theoretical disturbance at $(R,\Omega, t)$ 
resulting from a unit impulse originating at $(\rf, t_0)$ and
$\tilde \mathcal{G}_-$ is the time-reverse counterpart \citep{lb2000b,lb2004}. 

In the quiet Sun, the wave field can be well represented by a superposition of 
gravito-acoustic Green's functions of a standard reference model and the ingression and
egression functions are very similar. 
However, when the focus is located in an area of high magnetic activity, the measured signal 
deviates from that, effectively introducing a phase shift between the waves going in and out. 
This can be detected by computing the correlation that in the Fourier domain becomes the product of the Fourier transforms, 
$H_+ (\rf, \omega)$ and $H_-^* (\rf,\omega)$,
\begin{equation}
C(\rf) \equiv \int _{\omega _1}^{\omega _2}
 H_+(\rf, \omega) H_-^*(\rf,\omega) \rd \omega
\, .
\label{eq:C}
\end{equation}
The phase of the correlation, $\phi (\rf) \equiv \arg C(\rf)$, is related to the perturbed 
travel time by 
$\phi \simeq \omega \Delta\tau$. The seismic-holography maps are representations of
$\phi(\rf)$ as a measurement of the disturbance of the wave field due to the presence 
of magnetic regions.

Up to the present, far-side seismic holography has used  Green's functions calculated 
using the (acoustic) ray theory, e.g. \citet{lb2000b}. Formally, this 
approximation is only valid for  high frequencies and high angular degrees. 
In this paper, we develop the formalism
to determine the Green's functions through their polar expansion. Although the
method allows a full numerical computation both in the radial and angular 
variables, in the present work we have used the asymptotic approximation for 
the radial dependencies of the Green's function. On the other hand, the angular factors are
computed without any asymptotic approximation and hence the Green's functions
are expected to be more accurate for low and intermediate degrees. 

The outline of the paper is as follow.
In section 2 we describe the basic wave equations. Section 3 presents the calculation of the 
Green's functions using the gravito-acoustic ray-path approximation and section 4 
the Green's functions using the spherical polar expansion. 
The phase shift introduced by the non-adiabatic sub-surface layers of the 
Sun is reviewed in section 5. In section 6, we present an example of the improvement 
achieved by using the new Green's functions to calculate far-side maps of solar activity.

\section{Basic equations}

We start the analysis with the adiabatic wave equation in the Cowling and local
plane-parallel approximations. In terms of the scalar field 
$\tilde \Psi (t,\rr)= - \rho^{-1/2} \delta p$, 
where $\rho$ is the density of the background state and $\delta p (t, \rr)$ the 
Lagrangian pressure fluctuation, this equation is given by ---see \citet{g1993}---
\begin{equation}
c^{-2} \left( \frac{\partial^2}{\partial t^2} + \omega_c^2 \right)
\frac{\partial^2 \tilde \Psi}{\partial t^2} -
\frac{\partial^2}{\partial t^2} \nabla^2 \tilde \Psi - N^2 \nabla_{\rh}^2
\tilde \Psi = 0
\label{eqt}
\end{equation}
where $c$ and $N$ are the adiabatic sound speed and buoyancy frequency, defined as usual,
$\nabla_{\rh}^2$ represent the terms involving horizontal derivatives in
the $\nabla^2$ operator and the critical acoustic frequency is given by
\begin{equation}
\omega_c = \frac{c}{2H} \left( 1 - 2 \trr \cdot \nabla H \right)^{1/2}
\, .
\end{equation}
Here $H$ is the density-scale height and $\trr$ is an upward directed unit vector.

If the background state does not depend on time, as we will assume, the
scalar field can be Fourier decomposed such that the wave field is a
superposition of monochromatic waves given by
$\tilde \Psi (\rr, t) = \Psi (\rr, \omega) \exp{(-i\omega t)}$ and Eq.~(\ref{eqt}) 
becomes
\begin{equation}
\nabla^2 \Psi - \frac{N^2}{\omega^2} \nabla_{\rm h}^2 \Psi +
\frac{1}{c^2} \left( \omega^2 - \omega_c^2 \right) \Psi = 0
\, .
\label{eqw}
\end{equation}
We define the operator
\begin{equation}
\mathcal{L}_0 = \nabla^2 - \frac{N_0^2}{\omega^2} \nabla_{\rm h}^2 +
\frac{1}{c_0^2} \left( \omega^2 - \omega_{c0}^2 \right)
\, ,
\label{eql0}
\end{equation}
where $N_0$, $c_0$ and $\omega_{c0}$ are the buoyancy frequency, sound speed and cut-off 
frequency of a spherically-symmetric reference model. In this work we will use Model S
from \cite{j1996}. 
We assume that the actual solar oscillations satisfy a wave equation of the form 
\begin{equation}
\mathcal{L} \Psi = \mathcal{L}_0 \Psi + \mathcal{L}' \Psi = 0
\, ,
\label{eql}
\end{equation}
with $\mathcal{L}'\Psi$ accounting for all departures from Eq. (\ref{eqw}) (e.g. non 
adiabatic effects) and from the reference model 
(e.g. non spherical perturbations to the sound speed). 

\subsection{Integral representation}

Let us take a target point $\rf$, the focus, anywhere in the Sun. 
We introduce Green's functions, $G(\rr | \rf; \omega )$, that satisfy the 
source-point wave equation for the standard solar model,
\begin{equation}
\mathcal{L}_0 G(\rr| \rf; \omega)= - \delta(\rr - \rf)
\, .
\label{eqg}
\end{equation}
Multiplying Eq.~(\ref{eql}) by $G$, Eq.(\ref{eqg}) by $\Psi$, subtracting both,
integrating over a volume $\mathcal{V}$ that containing the focus
$\rf$ and, finally, using the Green theorem, the above equation can be written as
\begin{equation}
\int_{S} \left[ G {\bf V} \Psi - \Psi {\bf V} G \right] \cdot \rd {\bf S}
= \Psi (\rf,\omega) + \int_{\mathcal{V}} G \mathcal{L}' \Psi \rd \mathcal{V}
\, ,
\end{equation}
where $S$ is the surface enclosing $\mathcal{V}$, $\rd {\bf S}$ is directed outwards and 
$ {\bf V} = \nabla - (N_0^2/\omega^2) \nabla_{\rh}\,$.
If $S$ is the solar surface with radius $R$, the expression is simplified to
\begin{equation}
R^2 \oint_{S} 
\left( G  \frac{\partial \Psi}{\partial r} - \Psi \frac{\partial G}{\partial r} \right) 
\rd \Omega
= \Psi (\rf, \omega) + \int_{\mathcal{V}} G \mathcal{L}' \Psi \rd \mathcal{V}
\, .
\label{eqk1}
\end{equation}
A filtered version of the left-hand side of Eq.~(\ref{eqk1}) can be obtained 
from the observed wavefield and a reference model.
Moreover, if we compare the left-hand side of Eq.~(\ref{eqk1}) obtained from two sets of 
Green's functions, the differences will be due to the term with $\mathcal{L}'$. This includes
what can be called the ``anomalies'' (e.g. magnetic activity), but also systematic errors in 
the wave operator $\mathcal{L}_0$.

As indicated in the introduction we will consider Green's functions
$G_{\pm}$ corresponding to outgoing and ingoing waves from the focus. 
In this case, Eq.(\ref{eqk1}) is a Kirchhoff representation for the wave field $\Psi$ in the 
frequency domain. We compute the egression and ingression functions as
\begin{equation}
H_{\pm}(\rf, \omega) = \int_{\mathcal{P}_{\pm}} \Psiow
\frac{\partial G_{\pm}}{\partial r} (\Omega,\omega; \rf)
\rd \Omega
\; ,
\label{egres}
\end{equation}
where $\Psiow$  is the observed signal (Doppler shift velocity
in the frequency domain) and the integral is over a limited region of the solar surface, the pupil ($\mathcal{P}_{\pm}$). 
Correlating the ingression and egression functions one can 
expect to obtain a filtered version of the second term on the right-hand side of 
Eq.({\ref{eqk1}).

\section{Gravito-acoustic ray approximation}

We will compare our results with those obtained from the ray theory. Rather than
the acoustic approximation used by \citet{lb2000b},
we will consider a more general approximation.

For high frequencies, the general solution of $\mathcal{L}_0 \Psi =0$ 
can be expressed as a superposition of ``rays'' of the form 
\begin{equation}
\Psi (\rr) = A(\rr) \exp{\left( \pm i\int_l  k \, \rd l \right)}
\label{eqr}
\end{equation}
where we have introduced the local wave number, $k$. 
The integration is along a ray path and we have explicity indicated the two possible 
directions for each path. 
In this equation and
hereafter we will removed the subindex 0 for the reference model.

In the ray theory,
the ray path and the local wave vector, $\kk (\rr)=k_r \trr + {\bf k}_{\rm h}$,
are determined from a first order 3D Liouville-Green expansion (a ``WKB asymptotic 
approximation''). 
Specifically the dispersion relation is found to be given by
\begin{equation}
\frac{1}{c^2} \left(\omega^2 -\omega_c^2 \right) - k^2
+ \frac{N^2}{\omega^2} k_{\rm h}^2 = 0
\, .
\label{edisp}
\end{equation}
If the background state is spherically symmetric,
$k_{\rm h}^2 = L^2/r^2$, where $L$ is a constant.

The ray path is determined from the group velocity.
For the spherically symmetric case the rays are contained in a plane with a path given by 
\begin{equation}
\frac{\rd \theta}{\rd r} = \frac{v_{\theta}}{r v_r} 
\; .
\end{equation}
Here $r$ and $\theta$ are the usual polar coordinates and
the group velocity ${\bf v} = v_r \trr + {\bf v}_{\theta} \hat \theta$ has components
\begin{equation}
v_r = \frac{\partial \omega}{\partial k_r} = \frac{k_r \omega^3 c^2}
{\omega^4 - k_h^2 c^2 N^2}
\end{equation}
\begin{equation}
v_{\theta} = \frac{\partial \omega}{\partial k_{\rm h}} = 
k_{\rm h} \, \omega \, c^2 
\left( \frac{\omega^2 - N^2}{\omega^4 -k_{\rm h}^2 c^2 N^2} \right)
\, .
\end{equation}

The complex amplitude $A(r)$ is obtained from a second order Liouville-Green expansion, 
called the transport equation. In our case this equation reduces to  
\begin{equation}
\nabla \cdot \left( A^2 \left\{ \kk - \frac{N^2}{\omega^2} \kk_{\rm h} 
\right\} \right) =0
\label{etras}
\, .
\end{equation}
Integrating this equation over the volume of a ray tube, 
applying the Gauss theorem and taking $\Delta S \to 0$ for the tube section, one gets
\begin{equation}
A^2 \left( \kk - \frac{N^2}{\omega^2} \kk_{\rm h} 
\right) \Delta S \cdot \nn \simeq constant
\, ,
\end{equation}
where $\nn$ is a unit vector normal to the section. 
A representation of a ray tube similar to that considered here can be found in Figure 7 
of \cite{lb2000b}.
If we take the section such that $\nn$ is in the radial direction, the amplitude
can be determined by 
\begin{equation}
A \, r \left(k_r \sin \theta \, \Delta \theta \right)^{1/2}= constant
\, ,
\label{eamp}
\end{equation}
where $\Delta \theta$ is the angular size of the tube of section $\Delta S \to 0$.

\subsection{Green's functions}

The Green's functions defined by Eq. (\ref{eqg}) have also a 
solution of the form (\ref{eqr}) 
but with prescribed initial conditions, corresponding to a Dirac-delta impulse at the focus. 
Thus, the (complex) constant in Eq. (\ref{eamp}) is fixed. Specifically, 
the amplitude for a Green's function at any point $\rr$ far from the focus is given by
\begin{equation}
A_{\pm}(\rr |\rf) =  
\frac{\exp{(\pm i k_0 |\rr - \rr_{\epsilon}|)}}
{4\pi |\rr_{\epsilon} - \rf |} 
\frac{r_{\epsilon}}{r} \left| 
\frac{(k_r \sin \theta)_{\rr_{\epsilon}}}{k_r \sin \theta}  
\right|^{1/2}  
\left| \frac{\Delta \theta_{\epsilon}}{\Delta \theta} 
\right|^{1/2} 
\, .
\label{egray}
\end{equation}
Here $\rr_{\epsilon}$ is a point close to $\rf$ where the integration along the
ray path in (\ref{eqr}) starts up. For the Green's functions, here and in (\ref{eqr}), the 
$+$ sign corresponds to waves moving out of the focus $\rr_0$ (amplitude $A_+$) 
and the $-$ sign to waves moving into the focus (amplitude $A_-$). 

We are interested in computing outward and inward Green's functions but, to be more general, 
after a given number of bounces at the surface and the interior.
At the turning points, $k_r=0$ and the amplitude cannot 
longer be computed by means of Eq.(\ref{eamp}). In fact, we know that at the turning points
$A(\rr)$ is just retarded by $\pi/2$. 
Since these Green's functions are defined only within the resonance cavity, both
$\rr$ and $\rf$ must be inside the cavity. In particular, at least 
formally, the Green's functions cannot be computed in the photosphere or above.
Thus for a ray that initially goes inwards and then it is observed near the 
surface at its first arrival, the phase shift is $-\pi/2$. 
For any additional pair of inner and outer reflections the phase shift is $-\pi$. So after $s$ inner and $s-1$ surface bounces (hereafter $s$ skips), 
the phase shift is $-(2s-1)\pi/2$ and the inward and outward Green's functions at 
a point $\rr$ are given by 
\begin{eqnarray}
G_{\pm}(\rr |\rf ) = A_{\pm}(\rr |\rf )
\exp{ \left\{ \mp i \frac{(2s-1)\pi}{2} \right\} } & \times
\nonumber \\
\exp{ \left(\pm i \left\{ \int_{\rf \to \rr_1} k \rd l + 
2 (s-1) \int_{\rr_1 \to \rr_2} k \rd l
+  \int_{\rr_1 \to \rr} k \rd l \right\} \right) }
&
\label{egray2}
\end{eqnarray}
where $\rr_1$ and $\rr_2$ are the inner and upper turning points
and $A_{\pm}(\rr |\rf )$ is given by (\ref{egray}). The limits on the integrals 
indicate the interval over the ray path to be considered.

\section{Spherical polar expansion: Asymptotic approximation}

For a spherically symmetric reference model, 
Eq. (\ref{eqg}) can be solved by expanding the solution in terms of spherical harmonics in the form 
\begin{eqnarray}
G (\rr | \rf ) = \sum_{\ell=0}^{\infty} g^{\ell} (r | r_0)
\sum_{m=-\ell}^{\ell} Y_{\ell m}^* (\Omega_0)  Y_{\ell m} (\Omega)
= \nonumber \\
\frac{1}{4\pi} \sum_{\ell=0}^{\infty} g^{\ell} (r | r_0) (2\ell +1) P_{\ell}(\mu)
\label{edesc}
\end{eqnarray}
where $P_{\ell}(\mu)$ is a Legendre polynomial and 
$ \mu = \cos \theta \cos \theta_0 + \sin\theta \sin\theta_0 \cos(\varphi -
\varphi_0)$. Here $r\,$,$\theta$ and $\varphi$ are standard spherical coordinates and
$\Omega=(\theta,\varphi)$. 

Since we are considering a spherically symmetric reference model, we can choose,
without losing generality, the polar axis in a such a way that $\theta_0=0$ and 
$\varphi_0=0$ for the focus. In this way $\mu=\cos\theta$ and the Green's functions have not
dependence on $\varphi$. 

Inserting the decomposition given by (\ref{edesc}) into Eq. (\ref{eqg}), it is found that the
radial factors $g_{ell}$ satisfy the equation
\begin{equation}
\frac{\rd^2 \tilde g_{\ell}}{\rd r} + k_r^2 \tilde g_{\ell} =  
0 \qquad {\rm at} \quad \rr \neq \rf
\label{eag1}
\end{equation}
where $\tilde g_{\ell}= r g_{\ell}$ and $k_r$ is given by Eq. (\ref{edisp}) 
with $L^2=l(l+1)$. The Dirac-delta term is transformed into the following inhomogeneous 
initial conditions:
\begin{equation}
\tilde g_{\ell} (r_0^- |r_0)= \tilde g_{\ell} (r_0^+| r_0)
\label{eini1}	
\end{equation}
and
\begin{equation}
\frac{\rd \tilde g_{\ell}}{\rd r} (r_0^+| r_0)- 
\frac{\rd \tilde g_{\ell}}{\rd r} (r_0^- |r_0)= -\frac{1}{r_0}
\; .
\label{eini2}	
\end{equation}

We also recall that the angular factors, given by the Legendre polynomials, satisfies the 
equation
\begin{equation}
\frac{\rd^2 Q_{\ell}}{\rd\theta^2} + \kappa_{\ell}^2 Q_{\ell} =  0
\label{eag2}
\end{equation}
where $P_{\ell} (\cos \theta)= (\csc^{1/2} \theta) Q_{\ell} (\theta)$ and 
\begin{equation}
\kappa_{\ell}^2  = \left( \ell + \frac{1}{2} \right)^2 + \frac{1}{4} 
\frac{1}{\sin^2\theta} 
\, .
\end{equation}

\subsection{Asymptotic inwards and outwards Green's functions}

We are interested in keeping the same kind of ingoing and outgoing Green's functions than in 
the ray theory, since this will allow us to do a straightforward comparison.
Since we have done a separation of radial and angular variables, we need to search for inwards and 
outwards radial solutions and prograde and retrograde angular solutions and combine both 
in a suitable way. This kind of ``travel-wave'' solutions are well defined in the 
asymptotic limit and we start with that. 

In a first order Liouville-Green expansion the general solution 
of the angular factors can be written as
$Q_{\ell} (\theta) = \beta_{\ell +} B_{\ell +}(\theta) + \beta_{\ell -}  B_{\ell -}(\theta)$, where 
\begin{equation}
B_{\ell\pm} (\theta) = \kappa_{\ell}^{-1/2} \exp{\left(\pm i \int_{\theta_0}^{\theta} \kappa_{\ell} \, \rd 
\theta \right)}
\label{eqw2}
\end{equation}
and $\beta_{\ell +}$ and $\beta_{\ell -}$ are constants. 
The value $\theta_0$ will be taken at the focus.
In a similar way the solution of the radial factors can be written as 
$ \tilde g_{\ell} (r) = \alpha_{\ell +} C_{\ell +}(r) + \alpha_{\ell -} C_{\ell -}(r)$
where $\alpha_{\ell +}$ and $\alpha_{\ell -}$ are constants and
\begin{equation}
C_{\ell\pm} (r)= k_r^{-1/2} \exp{\left(\pm i \int_{r_1}^r k_r \rd r \right)}
\, .
\label{eqw1}
\end{equation}
Here $r_1$ is the inner turning point. 
The constants $\alpha_{\ell +}$ and $\alpha_{\ell -}$ 
can be obtained by introducing the equation 
$ \tilde g_{\ell} (r) = \alpha_{\ell +} C_{\ell +}(r) + \alpha_{\ell -} C_{\ell -}(r)$
into the initial conditions, (\ref{eini1}) and (\ref{eini2}). One gets
\begin{equation}
\alpha_{\ell\pm} = \frac{i}{2r_0} \frac{1}{\sqrt{k_r(r_0)}} 
\exp{\left(\mp i \int_{r_1}^{r_0} k_r \rd r \right)} 
\; .
\label{ealpha}
\end{equation}

In principle, we can take the outward prograde part of the solution, proportional to
$\alpha_{\ell +} C_{\ell +}(r) \beta_{\ell +} B_{\ell +} (\theta)$, as the Green's function going out
of the focus. 
However, the solution $\alpha_{·\ell +}C_{\ell+}$ corresponds to an outgoing wave emanating from the 
focus $r_0$ and reaching the surface without any reflection. Since
we are interested in the outgoing Green's function after $s$ skips, we need to define new inwards and outwards asymptotic solutions. 
Let us call $C_{\ell +}^s$ to the wave emanating inwards from the focus and now going outwards 
after $s$ skips. This function and his inwards counterparts are:
\begin{equation}
C_{\ell\pm}^s = C_{\ell\pm} 
\exp{ \left[
\pm i \left\{ 2 (s-1) \left( \int_{r_1}^{r_2} k_r \rd r - \frac{\pi}{2} \right)
-\frac{\pi}{2} \right\} \right] }	
\label{eqw1b}
\; ,
\end{equation}
where we have taken into account a phase jump of $-\pi/2$ every time the wave reach a 
turning point.

The angular factors has no turning points, so we can use the asymptotic functions defined 
in Eq. (\ref{eqw2}). Then, the Green's function for a given degree $l$ and frequency $\omega$
going out of the focus, proportional to $\alpha_{\ell +} C_{\ell +}^s(r) \beta_{\ell +} B_{\ell +} (\theta)$, is
\begin{eqnarray}
G_ {\ell +}^s (\vec r | \vec r_0, \omega) = \frac{i(2\ell+1)}{8 \pi r_0 \sqrt{k_r (r_0})} 
r^{-1} k_r^{-1/2} \kappa_{\ell}^{-1/2} \csc^{1/2}\theta \times
\nonumber \\
\exp{ \left\{ i \left( \int^{\theta}_{\theta_0} \kappa_l \rd \theta
+ \int_{r_0}^{r_1} k_r \rd r \right) \right\} } \times
\nonumber \\
\exp{ \left\{ i \left( 2 (s-1) \left( \int_{r_1}^{r_2} k_r \rd r - \frac{\pi}{2} \right)
- \frac{\pi}{2}
+ \int_{r_1}^r k_r \rd r \right) \right\} }
\, .
\label{eg0}
\end{eqnarray}
The ingoing Green's function is the complex conjugate, 
$G_{\ell -}^s = G_{\ell +}^{s*} \,$.

\subsection{Non-asymptotic inwards and outwards Green's functions}

Let us start with the angular factors. This corresponds to Eq. (\ref{eag2}) that of course has
the solution $Q_{\ell}=\sqrt{\sin\theta} P_{\ell}(\cos\theta)$.
To split it in prograde and retrograde parts we written the solution as 
\begin{equation}
Q_{\ell} (\theta) = \beta_{\ell +}(\theta) B_{\ell +}(\theta) + \beta_{\ell -}(\theta)  B_{\ell -}(\theta)
\; , 
\label{egq}
\end{equation}
where $B_{\ell\pm}$ are still given by Eq. (\ref{eqw2}).
In the asymptotic approximation $\beta_{\ell\pm}$ were taken to be constants but
now they are functions that change smoothly, at least where the asymptotic approximation is 
expected to work. Thus we replace the constants by osculating parameters that satisfies 
the additional condition
\begin{equation}
Q_{\ell}' (\theta)= \beta_{\ell +} (\theta) B_{\ell +}'(\theta) + \beta_{\ell -}(\theta) B_{\ell -}' (\theta)
\label{eqb}
\end{equation}
where the primes means derivatives respect to $\theta$. With this definition it is follow 
that
\begin{equation}
\beta_{\ell +} = \frac{i}{2} ( Q_{\ell} B_{\ell -}' - Q_{\ell}' B_{\ell -})
\quad ; \qquad 
\beta_{\ell -} = -\frac{i}{2} ( Q_{\ell} B_{\ell +}' - Q_{\ell}' B_{\ell +})
\end{equation}
Since $Q_{\ell}$ is known, $\beta_{\ell\pm}$ and hence the prograde $\beta_{\ell +} B_{\ell +}$ and 
retrograde parts $\beta_{\ell -} B_{\ell -}$ are obtained.

For reasons explained later, for the applications shown in this work we will use the asymptotic approximation for the radial factors. 
However, for completeness, we explain here how to obtain the non-asymptotic radial Green's functions. 
The general solution of Eq.(\ref{eag1}) can be written as 
\begin{equation}
\tilde g_{\ell} (r|r_0) = \alpha_{\ell +} (r) C_{\ell +}^s(r) + \alpha_{\ell -} (r) C_{\ell -}^s(r)
\label{egg}
\end{equation}
where $C_{\ell\pm}^s$ are given by (\ref{eqw1b}). 
Here again, to fix the osculating parameters we add the condition 
\begin{equation}
\tilde g_{\ell}' (r|r_0)= \alpha_{\ell +} (r) (C_{\ell +}^s)'(r) + \alpha_{\ell -}(r) (C_{\ell -}^s)' (r)
\end{equation}
from which it is found that
\begin{equation}
\alpha_{\ell +} = \frac{i}{2} ( \tilde g_{\ell} (C_{\ell -}^s)' - \tilde g_{\ell}' C_{\ell -}^s)
\quad ; \qquad 
\alpha_{\ell -} = -\frac{i}{2} ( \tilde g_{\ell} (C_{\ell +}^s)' - \tilde g_{\ell}' C_{\ell +}^s)
\, .
\label{eqd}
\end{equation}
Thus, for a given solution $\tilde g_{\ell}(r|r_0)$,
the functions $\alpha_{\ell +}(r)$ and $\alpha_{\ell -}(r)$ can be computed. 
Note that $\tilde g_{\ell}(r|r_0)$] satisfies Eq. (\ref{eag1}) and since we are dealing with waves 
reflected in the interior, the solution must be regular at the center. 
With this condition alone, $\tilde g_{\ell}(r|r_0)$ is determined up to a constant factor. 
On the other hand $\tilde g_{\ell} (r_0|r_0)$ can be taken as the one given by the asymptotic 
solution, namely 
$\tilde g_{\ell} (r_0)= \alpha_{\ell -}(r_0) C_{\ell -}^s(r_0) + \alpha_{\ell +}(r_0) C_{\ell +}^s(r_0)\,$, with
$\alpha_{\ell\pm}(r_0)$ given by Eq.(\ref{ealpha}) and $C_{\ell\pm}^m$ by Eq.(\ref{eqw1b}).
This fixes $\tilde g_{\ell}(r|r_0)$ completely and hence the inward and outwards 
solutions, $\alpha_{\ell -}(r) C_{\ell -}^s(r)$ and $\alpha_{\ell +}(r) C_{\ell +}^s(r)$ respectively.
Combining the outward solution with the prograde solution of the angular part, the outgoing 
wave is given by
\begin{eqnarray}
G_{\ell +}^s (\vec r|\vec r_0) = -\frac{2l+1}{16\pi r}\csc^{1/2} \theta \; \times
\nonumber \\
\left\{ Q_{\ell }(\theta) B_{\ell -}'(\theta) - Q_{\ell}'(\theta) B_{\ell -}(\theta) \right\} B_{\ell +}(\theta) 
\; \times
\nonumber \\
\left\{ \tilde g_{\ell}(r) C_{\ell -}'(r) - \tilde g_{\ell}'(r) C_{\ell -}(r) \right\} C_{\ell +} (r)
\, .
\label{ego}
\end{eqnarray}
The ingoing Green's function is the complex conjugate.
$G_{\ell -}^s = G_{\ell +}^{s*} \,$.

\section{The Phase Shift}

\begin{figure}
\plotone{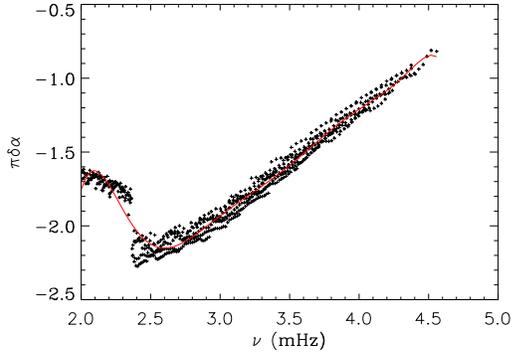}
\caption{Phase shift computed using equation (\ref{eshift}). Eigenfrequencies, with
degrees $\ell=20-80$, are taken from MDI data. Model S of
Christensen-Dalsgaard (1996) have been used. The solid thick line represent a
fitted polynomial of degree 10.}
\label{fshift}
\end{figure}

In the uppermost layers of the Sun, the asymptotic approximation fails.
Moreover, non adiabatic effects and the dynamics of convection, that are not included in
the wave equations, are too large in these layers to be neglected.
These effects were taking into account by \citet{lb2004} by using 
empirical control correlation functions.
Here we use a completely different approach.

When computing the wave functions in the asymptotic approximation, these effects can be 
taken into account by introducing an additional phase shift at the upper turning 
point ---e.g. \citet{g1984}, \citet{dg1984}---. 
In the same way the Green's functions can be modified to include such effects,
namely
\begin{equation}
\mathcal{G}^s_{\ell}(r,\theta|r_0 ,\theta_0) = G_{\ell}^s(r,\theta|r_0, \theta_0) 
\exp{(2i\tilde s\pi\delta\alpha)}
\label{egreens}
\end{equation}
where $2\pi\delta \alpha-\pi/2$ is the phase shift after one surface reflection. To a good approximation, $\delta \alpha$ is a function of 
frequency alone. 
If we were to include just the formal bounces at the surface, $\tilde s=s-1$. However the pupil and, depending on the problem, the focus 
are so close to the upper turning points that each of them can account for an extra shift in the phase of, say, a half bounce. 
Thus in practice one should take $\tilde s=s-1/2$ if the focus is below the uppermost layers or $\tilde s=s$ if the focus is too close to the surface.

As shown by \citet{cp1992}, a theoretical phase
shift at every point, $\delta \alpha(r)$, can be computed by comparing the asymptotic 
approximation and the numerical solutions of the wave equations. This would allow 
to use Eq.(\ref{egreens}) to correct the Green's functions at any point in the interior, but,
unfortunately, our tests have shown that this correction is not enough to
achieve good seismic images. The reason is that they did not include 
any non adiabatic or convective terms in the full wave equations, and this is not a simple 
task.

For this work, we have computed an $r$-independent phase shift based on the observed  
frequencies of the normal mode of oscillations. This way we include all the neglected
aspects in the wave equations but
the focus must be assumed to be above or below these uppermost layers. 
Moreover, the method is limited to the asymptotic approximation for
the radial factors of the solution. 
Specifically, $\delta\alpha(\omega)$ can be estimated by inserting the observed 
frequencies into the asymptotic expression
\begin{equation}
\int_{r_1}^{r_2} k_r \rd r = \pi \left( n- \frac{1}{2} - \delta\alpha \right) 
\,,
\end{equation}
where $n$ is an integer, the radial order of the mode. In our case the wave number is 
given by Eq. (\ref{edisp}) and hence
\begin{equation}
\int_{r_1}^{r_2} \left[ \frac{1}{c^2} \left(\omega^2 -\omega_c^2 \right) 
- \left( 1-\frac{N^2}{\omega^2} \right) \frac{L^2}{r^2} \right]^{1/2}
\rd r = \pi \left( n- \frac{1}{2} - \delta\alpha \right) 
\, .
\label{eshift}
\end{equation}

To obtain $\delta\alpha(\omega)$, eigenfrequencies from $\nu =2500\,\mu$Hz to 
$\nu =4500\,\mu$Hz and degrees $\ell=20-80$ 
were taken from MDI data. Modes of lower degrees are rejected because they do not satisfies
the Cowling approximation as good as intermediate and high-degree modes.
In addition, Model S from \citet{j1996} was used. A fitted 
polynomial of degree 10 to Eq. (\ref{eshift})
was considered. The result is presented in Fig. \ref{fshift}.
Notice that $\pi \delta \alpha$ changes by more than $1.5\,$rad in the frequency range of 
interest.

\section{Numerical results} 

To validate the Green's functions calculated using this new approach, we apply them to 
create seismic maps of active regions at the far side of the Sun. In the case of far-side 
seismic holography, the focus is located at (or just below) the surface of the far hemisphere
of the Sun and the pupil is an annulus surrounding the antipodes of the focus. Although 
seismic maps of the full far-side hemisphere can be calculated using waves following one, two and three-skip ray paths
before arriving at the pupil \citep{braun2001}, for simplicity, here we will concentrate only on the central part of 
the maps. Traditionally, this is calculated considering waves following a 2-skip ray path, as described by \citet{lb2000b}. 

\subsection{Comparison of ray theory and polar expansion}

As it has been done in previous work, we normalize the Green's functions at each frequency
to its mean absolute value.
Fig.~\ref{fgreen1} shows these normalized Green's functions against the polar angle
for different frequencies. Results for the polar expansion and the 
gravito-acoustic rays are compared. Eq. (\ref{ego}) has been used for the polar expansion, but 
with the asymptotic approximation for the radial factors,  while 
for the ray theory Eq. (\ref{egray2}) has been considered. In both cases the surface phase 
shift shown in Fig. \ref{fshift} has been added according to Eq. (\ref{egreens}).
Finally, what we actually show are the derivatives of the Green's functions that we will use
to compute the ingression and egression fields following Eq. (\ref{egres}). 
The focus is located at $r_0=0.9995R_{\odot}$, $\theta_0=0$ 
and the Green functions are shown at the upper turning point after two skips.

In the ray theory, for a given number of skips, and a given frequency $\omega$, each 
point in the pupil corresponds to a given 
value of the horizontal wave number at the focus, $k_h(r_0)$ and hence of 
$L=\ell (\ell+1)$. Here $\ell$ is a real constant. 
Our pupil goes from $\theta_{\mathrm{min}}=2$ to $\theta_{\mathrm{max}}=3\,$rad; this
corresponds to and interval in ``degrees'' from $l_{\mathrm{min}} \simeq 15$ to 
$l_{\mathrm{max}} \simeq 28$ for $\nu = 2500\,\mu$Hz and 
$l_{\mathrm{min}} \simeq 27$ to $l_{\mathrm{max}} \simeq 50$ for $\nu = 4500\,\mu$Hz.
In the polar expansion there is not such relation, in fact
in Fig. \ref{fgreen1} Green's functions for degrees up to $l=60$ have been added up,
the same for all the frequencies.

As it can be seen in Fig. \ref{fgreen1},
there is a systematic shift between the Green's functions computed with the two different
approaches, but the solutions are closer at low angles; in fact in the ray theory these
angles correspond to higher $l$ where this approximation is expected to work better. 
On the other hand
the solutions of the polar expansion approximation look more irregular at low frequencies; 
this is related to the fact that at these frequencies the upper turning point $r_2$ 
is too close to the focus $r_0$.

\begin{figure}
\plotone{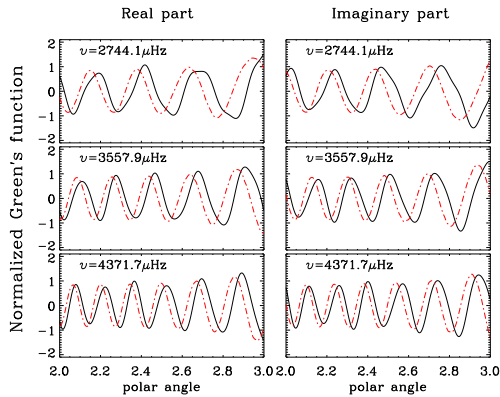}
\caption{Comparison of Green's functions at selected frequencies. The black 
continuous lines correspond to the Green's functions calculated using the 
Legendre polynomial decomposition and the red dashed ones to those calculated 
using a gravito-acoustic ray path approximation.}
\label{fgreen1}
\end{figure}

We use both sets of Green's functions described in this paper, as well as the traditional ones (acoustic ray approximation), 
to calculate maps of magnetic activity at the 
non-visible hemisphere of the Sun. Fig.~\ref{fmaps} presents far-side maps calculated using 
Global Oscillation Network Group (GONG, \cite{harvey1996}) data for September 1 and 2 2005 for the three sets of Green's functions. The maps 
clearly show the signature of active region NOAA-10808 as a large negative phase shift. The 
Green's functions used for the analysis include frequencies computed from 
$\nu = 2500\,\mu$Hz up to $4500\,\mu$Hz with a step of $\sim 8\,\mu$Hz. Angles go from 
$\theta =2$  to $3\,$rad with a step of $0.0016\,$rad. Active region NOAA 10808 appeared at 
the East limb of the Sun September 7 2005.

\begin{figure}
\plotone{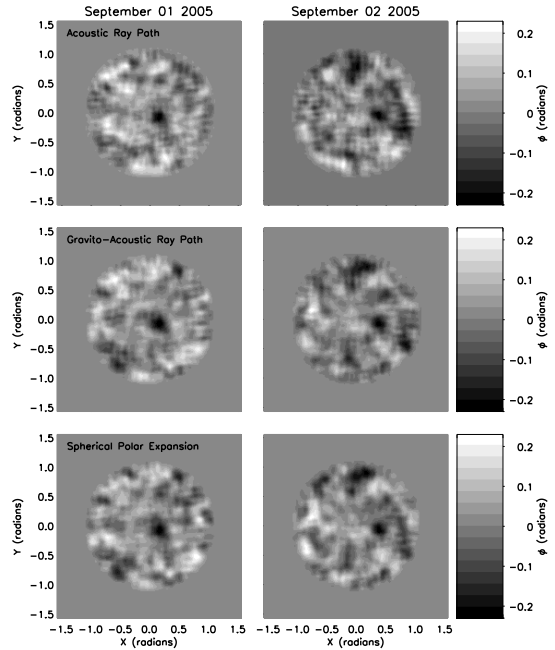}
\caption{Postel-projected maps of the far side calculated using 
the traditional acoustic ray approximation (top), the gravito-acoustic ray approximation (center) and the spherical polar expansion 
(bottom) Green's functions. The maps extend to approximately 60 degrees from 
the antipodes of the center of the solar disk facing the Earth. The areas of 
large negative phase shift ($\phi$) are related to areas of strong magnetic field. 
In this case, the strong signature corresponds to active region NOAA 10808 5-6 days before it appeared at the East limb of 
the Sun on September 7 2005.}
\label{fmaps}
\end{figure}

The signature of the active region is very similar in the maps for all the cases. However, 
while the noise level is similar in all three cases, the spatial distribution is only correlated between the maps 
calculated using the two types of Green's functions described in this 
paper. This may be to the different approaches used for the sub-surface phase 
shift correction, explained in section 5. However more research is necessary in order to understand the different contributors to the noise.

\subsection{Green's functions combining different number of bounces}

As mentioned before, in the polar expansion approximation there is not a one to one
relation between the angle in the pupil and the mode degree. Thus, even for the same
pupil and the same angular degrees it is possible to combine the Green's functions from 
different bounces.
In Fig. \ref{fgreen2} we show the Green's functions from the polar expansion
against the polar angle. In this case we have used degrees from $\ell=15$ to $\ell=100$.
The Green's functions are shown for $s=2$, $s=3$ and the sum of both (with the normalization 
at each frequency done after the sum).
We do not consider the $s=1$ case since most of the signal for the first bounce in our pupil 
comes from very low degrees for which the Cowling approximation does not work properly.
As it can be seen, the maximum
amplitude of these Green functions change both with $s$ and frequency. 

\begin{figure}
\centering
\plotone{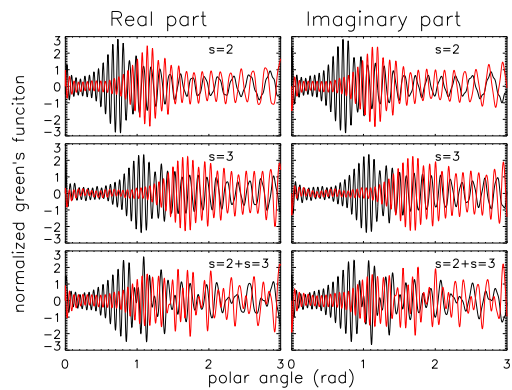}
\caption{Normalized Green's functions against polar angle from the focus computed with the 
polar expansion technique for 
$s=2$, $s=3$ and the sum of both. Black lines are for $\nu=3000\,\mu,$Hz and
the red ones for $\nu=4500\,\mu,$Hz.}
\label{fgreen2}
\end{figure}

We calculate a series of far-side maps using the polar-expansion Green's functions that 
combine two and three skips. In Fig. \ref{fmaps2}, we compare  these maps with those using 
Green's functions that include only $s=2$. A sequence of three days spanning November 27 
to 29 2006 showing a weaker active region just south of the equator an past the central meridian is
presented. This active region was at the same location in the front site the previous Carrington 
rotation (NOAA 10923) and reappeared at the East limb December 5 2006 (NOAA 10930). We find 
that the signature associated to the magnetic area is enhanced when using the combination of 
2 and 3 skips by approximately 20-30$\%$ compared to the 2 skips approach used 
for the center of the map.

\begin{figure}
\centering
\plotone{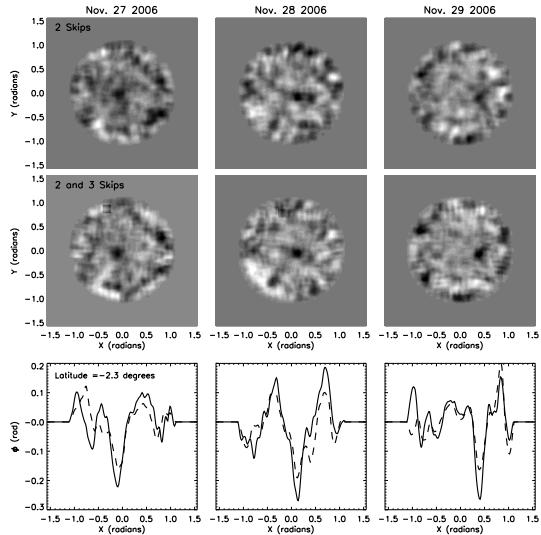}
\caption{Postel projected maps of the far side calculated using the spherical polar expansion
Green's functions for two skips (top) and a combination of two and three skips (center). 
The signature of active regions NOAA 10930 can be seen as at the far 
side hemisphere, it appeared at the front side on December 5 2006 just below the solar 
equator. The bottom panels show a cut of the far-side maps at heliocentric latitude -2.3 
degrees. The phase shift associated to the magnetic area is enhanced when using the 
combination of 2 and 3 skips by approximately 20-30$\%$ compared to the 2 skips 
approach used for the center of the map.}
\label{fmaps2}
\end{figure}

\section{Conclusions}

We have implemented  a holographic, or migration technique, that includes a factorization 
of the Green's functions in angular and radial parts. This technique is more general than 
the one provided by the ray theory. We show examples of how these new Green's 
functions can be used to improve far-side seismic images by combining the Green's functions 
corresponding to travel waves with two and three skips or inner reflections. 

\citet{zhao2007} and \citet{ilonidis2009}
using a time-distance technique have demonstrated that 
adding more skips to the calculation of far-side maps improves the signal-to-noise. 
However, in their analysis, they combine seismic images obtained from different skip ray paths after they are calculated.
Here, we show that by combining the Green's functions associated to two and three skips 
before calculating the map in seismic holography we improve the signal to noise in a
similar way. Since the spherical geometry 
of the observational pupil makes the calculation of the ingression and egression functions 
for far-side maps computationally expensive, the solution presented here 
is an ideal way of 
improving the signal to noise without the need for increasing the computation time.
Our approach is also formally different, since 
we sum the Green's functions from different bounces without a previous normalization.

There are some previous works where Green's functions beyond the ray approximation were 
used, for instance in \citet{lb2004}, but there a completely plano-parallel approximation
was considered and hence the technique was limited to focuses in the front-side and
close to the surface. 
On the other hand for our Green's functions, 
although we have shown here the application to the particular case of far-side seismic holography, 
the technique is general enough to be applied in the front side and for focuses under the surface, 
even below the convection
zone, at least formally. In that scenario, the signal-to-noise increases considerably and
more work would be required to test the inferences.

An interesting result of our work is the fact that the sub-surface phase shift, that has been 
recognized as a fundamental correction to any Green's function in the past, can be computed
by using the observed frequencies of the normal modes of oscillations. This opens a link 
between local and global helioseismology and perhaps the issues of the local approach can be 
used to learn something about the global one and vice versa. 

\acknowledgements
The authors thank C. Lindsey and D. Braun for their valuable comments on the paper and 
for making available large part of the far-side code used here.
This research was supported by grant PNAyA2007-62650 from the Spanish National Research Plan
and by NASA grant NAG 5-11703. SOHO
is a project of international cooperation between ESA and NASA.
This work utilizes data obtained by the GONG
program, managed by the National Solar Observatory, which is operated by AURA,
Inc. under a cooperative agreement with the National Science Foundation. The 
data were acquired by instruments operated by the Big Bear Solar Observatory, 
High Altitude Observatory, Learmonth Solar Observatory, Udaipur Solar 
Observatory, Instituto de Astrofísica de Canarias, and Cerro Tololo

\end{document}